\documentclass[doublecol]{epl2}

\title{Out of equilibrium dynamics in the bidimensional spin-ice model}

\author{Demian Levis\inst{1} \and Leticia F. Cugliandolo\inst{1}} 

\institute{
 \inst{1} Universit\'e Pierre et Marie Curie
  - Paris 6, Laboratoire de Physique Th\'eorique et Hautes Energies,
  4, Place Jussieu, Tour 13, 5\`eme \'etage, 75252 Paris Cedex 05,
  France
}

\pacs{74.40.Gh}{Fluctuation phenomena, nonequilibrium processes}
\pacs{75.10.Hk}{Ising model, magnetic ordering}
\pacs{75.40.Mg}{Computer modeling and simulation of magnetic critical points}

\abstract{
We study the  dynamics of $2d$ spin-ice following a quench
from a fully disordered initial condition (equilibrium at infinite temperature) 
into its disordered, ferromagnetic and
antiferromagnetic phases. We analyze the evolution of the density of 
topological defects and we show that these take finite density over very long
periods of time in all kinds of quenches.
We identify the leading mechanisms for the 
growth of domains in the ordered phases and we 
evaluate the (anisotropically) growing lengths involved in 
dynamic scaling.
}

\begin{document}

\maketitle

% Intro general. Frustracion y regla del hielo

In a large class of condensed-matter systems the tendency to local
ordering is hampered by constraints. Frustration entails,
typically, a non-vanishing entropy at zero temperature.

The prototypical example is {\it water ice} for which this zero point entropy has been 
measured in the 30s~\cite{Giauque1936}. Pauling explained this feature with 
a model in which the O atoms occupy the vertices of a coordination
four lattice and exactly two H atoms are near while the other two H atoms are shifted away
from each vertex~\cite{Pauling1935}. The large degeneracy of such locally electro-neutral 
ground states gives rise to the zero point entropy.

A residual entropy has also been measured
in frustrated magnets~\cite{Balents2010,DiepBookCH7} such
as~$\mbox{Ho}_2\mbox{Ti}_2\mbox{O}_7$~\cite{Harris1997}. In
these~\emph{spin-ice} samples, the relevant interacting degrees of
freedom are classical Ising spins located at the nodes of a pyrochlore
lattice and aligned with the local axis connecting two corner-sharing
tetrahedra~\cite{DiepBookCH7}.  Each tetrahedron can be seen as a
vertex in a $3d$ lattice taking one out of sixteen possible
configurations. The energy is locally minimized when two spins point
inward and two outward verifying the ice rule or zero-divergence  condition for the
magnetic moments carried by the spins. The other ten
`defects' have a magnetic charge $q=\vec {\nabla} \cdot \vec{S}$. 
Experimentally, the statistical weights of the vertices can be tuned by applying 
 uniaxial pressure that 
 break  the isotropy of the ice-model but preserve the $Z_2$ symmetry~\cite{Mito2007}. 
 Magnetic fields along different crystallographic axes can also be used to modify the 
 vertex weights.
 
Two dimensional spin-ice samples have also been produced in the laboratory. 
On the one hand, high magnetic fields project the system onto 
$2d$ Kagome slices~\cite{Fennell2007}; on the other hand, `artificial' spin-ice on a $2d$ 
square lattice can be manufactured with litography~\cite{Wang2006,Moller2006}. 
Although the magnetic moments  in artificial spin-ice are {\it a priori} athermal, an effective 
thermodynamics can be generated with applied rotating magnetic 
fields~\cite{Nisoli2007,Budrikis2010}.
%The dynamics induced by this procedure have been discussed in~\cite{Budrikis2010} 
%and in particular 
%the dynamics after the reversal of a strong external magnetic field (quench) ~\cite{Mellado2010}. 
%This procedure is used experimentally to find dynamical evidences for the presence of magnetic 
%monopoles and associated Dirac strings ~\cite{Mengotti2011}. 
More recently, thermal excitations have also been engineered in these 
samples~\cite{Kapaklis2011}.

In recent years, research in this field has been boosted by the exciting 
proposal that topological defects, in the form of magnetic monopoles and their attached Dirac strings,
could be observed in spin-ice samples~\cite{Castelnovo2008a,Jaubert2009,Castelnovo2010,Morris2009,Slobinsky2010,Mengotti2011}.
Following the proposal in~\cite{Castelnovo2008a} experiments have shown evidence for this kind of excitation~\cite{Morris2009,Slobinsky2010,Mengotti2011}. 
Reaction-diffusion arguments have been used to estimate
the density of defects in the disordered phase of $3d$ spin ice~\cite{Castelnovo2010}.
As far as we know no studies of dynamics in the ordered phases nor beyond this simple modeling
have been performed.

Here we choose a different approach to address the dynamics of spin-ice models. For the sake 
of simplicity we focus on thermal quenches in the  $2d$ square lattice spin ice model built as a stochastic
extension of the celebrated {\it 6 vertex model} of statistical mechanics~\cite{BaxterBook}.
We use a rejection-free
continuous-time Monte Carlo (MC) algorithm~\cite{Barkema-Newman_Book},
with local spin-flip updates and non-conserved
order parameter, that allows for thermally-activated creation of
defects. The longest time reached with this method, once translated in terms of usual MC sweeps, is of the order of
$10^{16}$ MCs, a scale practically unreachable with usual Metropolis algorithms. This allows us 
to identify the equilibrium phase diagram and to analyze different 
dynamic regimes. Our dynamic results are manifold. We reproduce known facts of the dynamics of 
spin ice samples, as the existence of long-lived metastable states although 
with no need of long-range interactions.  We prove that the dynamics
after a quench into the FM and AF phases conforms to the domain-growth scaling 
picture~\cite{Bray2002} and we identify the anisotropic growing lengths. We derive a large number of 
new results that, we propose, should 
be realized experimentally~\cite{Wills2000}. Our findings should also be of interest
to the integrable systems community.

\begin{figure}[h]
%CONFIGURACIONES DE VERTICES
\centering
\onefigure[scale=1.05]{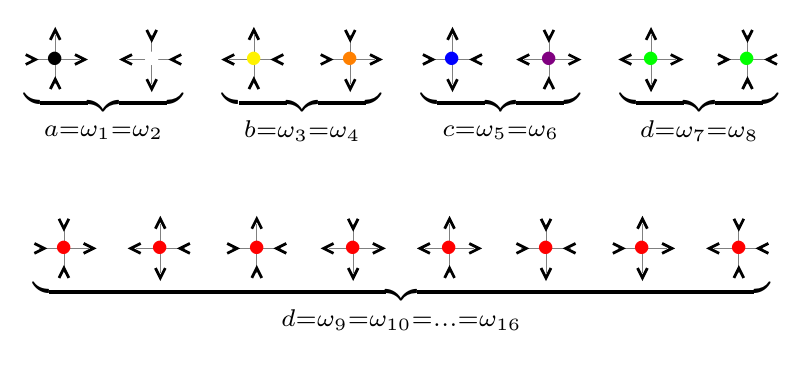} 
\caption{(Color online.)  The sixteen vertex configurations on the $2d$ square lattice 
and their  weights. The first six vertices verify the ice-rule.  The next pair
  completes the eight-vertex model and have charge $q=\pm 2$. The last eight vertices have charge $q=\pm 1$.
  This color code is used in
  Fig.~\ref{Evolution}.}
\label{vertex_configurations}
\end{figure}

{\it The model.}
%definicion del model 16-vertex que usamos
We consider an $L\times L$ square lattice with unit spacing and
periodic boundary conditions and we vary the linear size $L$ from 10 to 
300. Each edge is occupied by an arrow modeled as a binary
variable $S=\pm 1$.   We assign a Boltzmann weight $\omega_k
\propto e^{-\beta\epsilon_{k}}$ to each of the $k=1,\dots,2^{4}$
four-arrow vertex configurations. The energy is
%\begin{equation}%el Hamiltoniano
$ H=\sum_{k=1}^{16}\epsilon_{k}n_{k}$,
%\label{Hamiltonian16v}
%\end{equation}
where $n_{k}$ is the number of vertices of type $k$. We set
$\omega_1=\omega_2=a$, $\omega_3=\omega_4=b$, $\omega_5= \omega_6=c$
for the ice-rule vertices and $\omega_7=\omega_8=d$,
$\omega_9=...=\omega_{16}=d$ for the 2-fold and 1-fold defects,
respectively, ensuring invariance under reversal of all arrows 
(see Fig.~\ref{vertex_configurations}). Henceforth we measure the
weights in units of $c$.

\begin{figure}[h]
\centering
 \onefigure[scale=0.45,angle=-90]{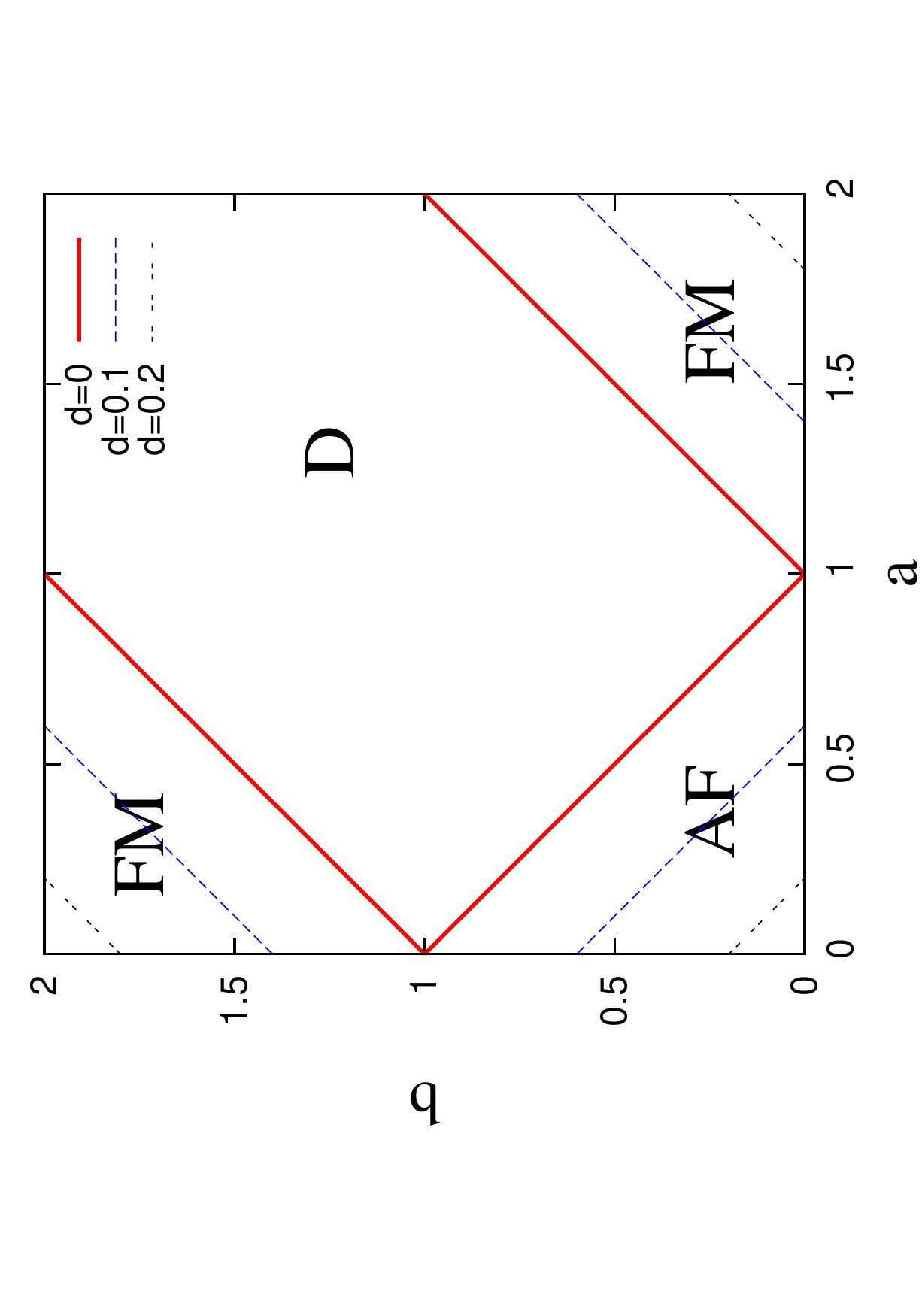}  
\par
\caption{(Color online.)  $(a,b)$-plane projection of the
  sixteen-vertex model phase diagram. The solid (red) lines are the
  six-vertex model first-order  transition between disordered
  (D) and  ferromagnetic (FM) phases, and 
 KT transition between disordered (D)  and antiferromagnetic (AF) phase.  
 The second-order transition lines for $d\neq 0$ are
  shown with dashed (blue and black) lines.}
\label{PhaseDiagram16}
\end{figure}

{\it Equilibrium properties.}
In the $2d$ six-vertex model defects are forbidden. A host of exact equilibrium properties of
this model 
were derived with the Bethe {\it Ansatz} and {\it via} mappings to random matrix theory, algebraic 
combinatorics (domino tilings) and crystal growth~\cite{BaxterBook}.  Depending on
the weight of the vertices the system sets into a quasi long-range
ordered paramagnetic (PM) or disordered (D), two ferromagnetic (FM) and one
anti-ferromagnetic (AF) phases separated by different transition lines. 
%A few studies of the out of equilibrium 
%dynamics of frustrated magnets in
%general~\cite{Wills2000}, using techniques and analysis pertaining to 
%spin-glasses, and spin-ice in 
%particular~\cite{Castelnovo2008a,Jaubert2009,Castelnovo2010,Slobinsky2010},
%with special interest in the study of topological defects,  appeared in the
%literature recently.
The four phases of the six-vertex model ($d=0$) are characterized by the parameter
%\begin{equation}%Delta 6V
 $\Delta_6=(a^{2}+b^{2}-1)/(2ab)$~\cite{BaxterBook}.
 (i) For $\Delta_6>1$ and $a>b+1$ the vertices of type 1 and 2 (type 3 and 4 for $b>a+1$) 
are statistically favored and the system is frozen into one of its two symmetric 
ground states with perfect FM order and no low energy excitations. At $|a_c^{FM}-b|=1$, 
i.e. $\Delta_6=1$, there are two first-order phase transition.  (ii) For $-1<\Delta_6<1$ the 
system is quasi long-range ordered. The ice-rule is strong enough to
prevent full de-correlation at any temperature.  At $a_c^{AF}=1-b$,
i.e. $\Delta_6=-1$ there is a Kosterlitz-Thouless (KT) phase
transition.  (iii) For $\Delta_6<-1$ vertices of type 5 and 6 are
favored and the system orders into an AF state populated by low energy
excitations. The phase diagram is shown in Fig.~\ref{PhaseDiagram16}
with solid (red) lines. This scenario is modified by the ice-rule breaking 
vertices ($d\neq 0$) as we discuss below.

%Los resultados de la simulacion: 1/2 p.
\begin{figure}[h]
\centering
 \includegraphics[scale=0.45,angle=-90]{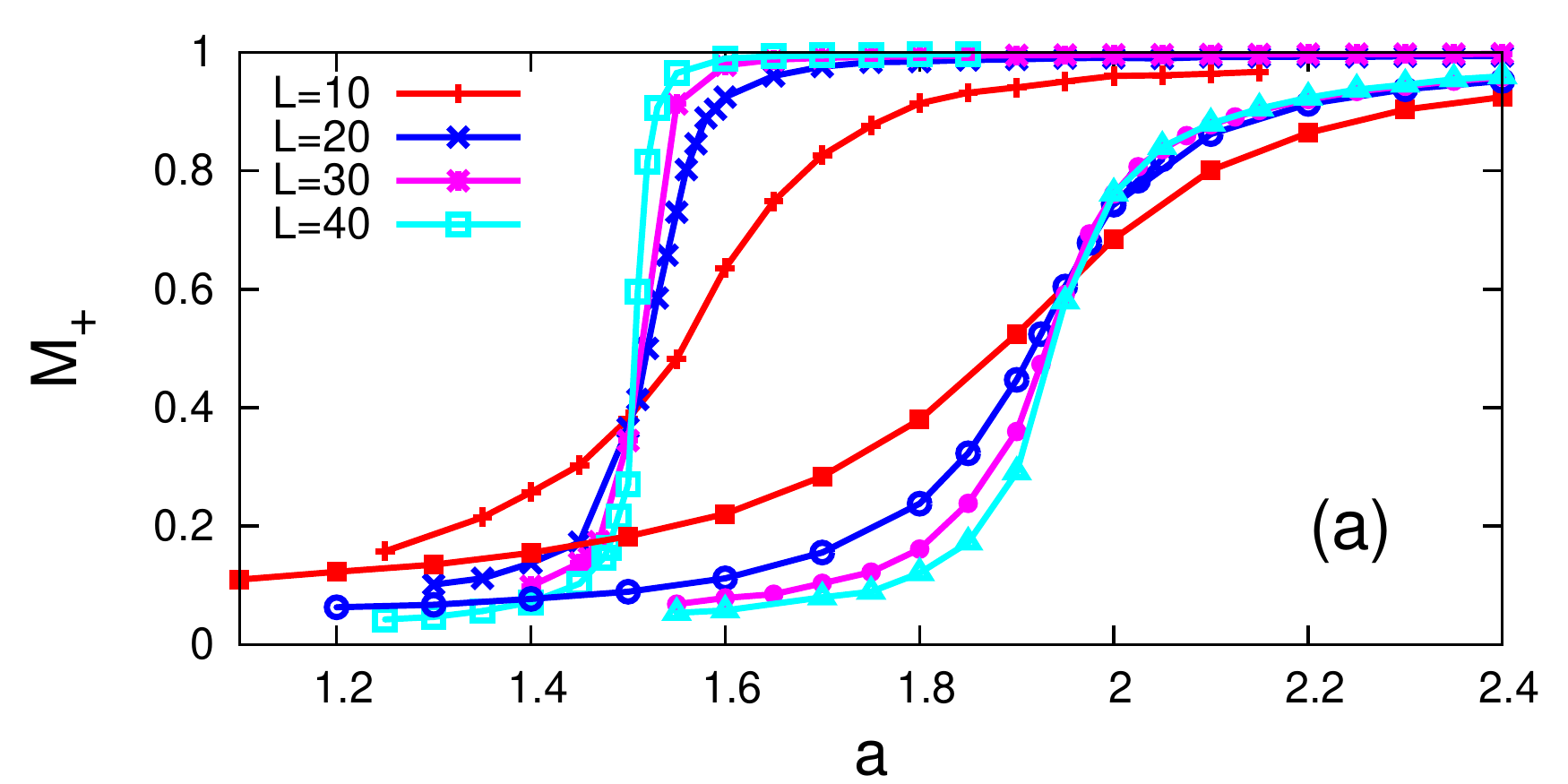} 
\includegraphics[scale=0.45,angle=-90]{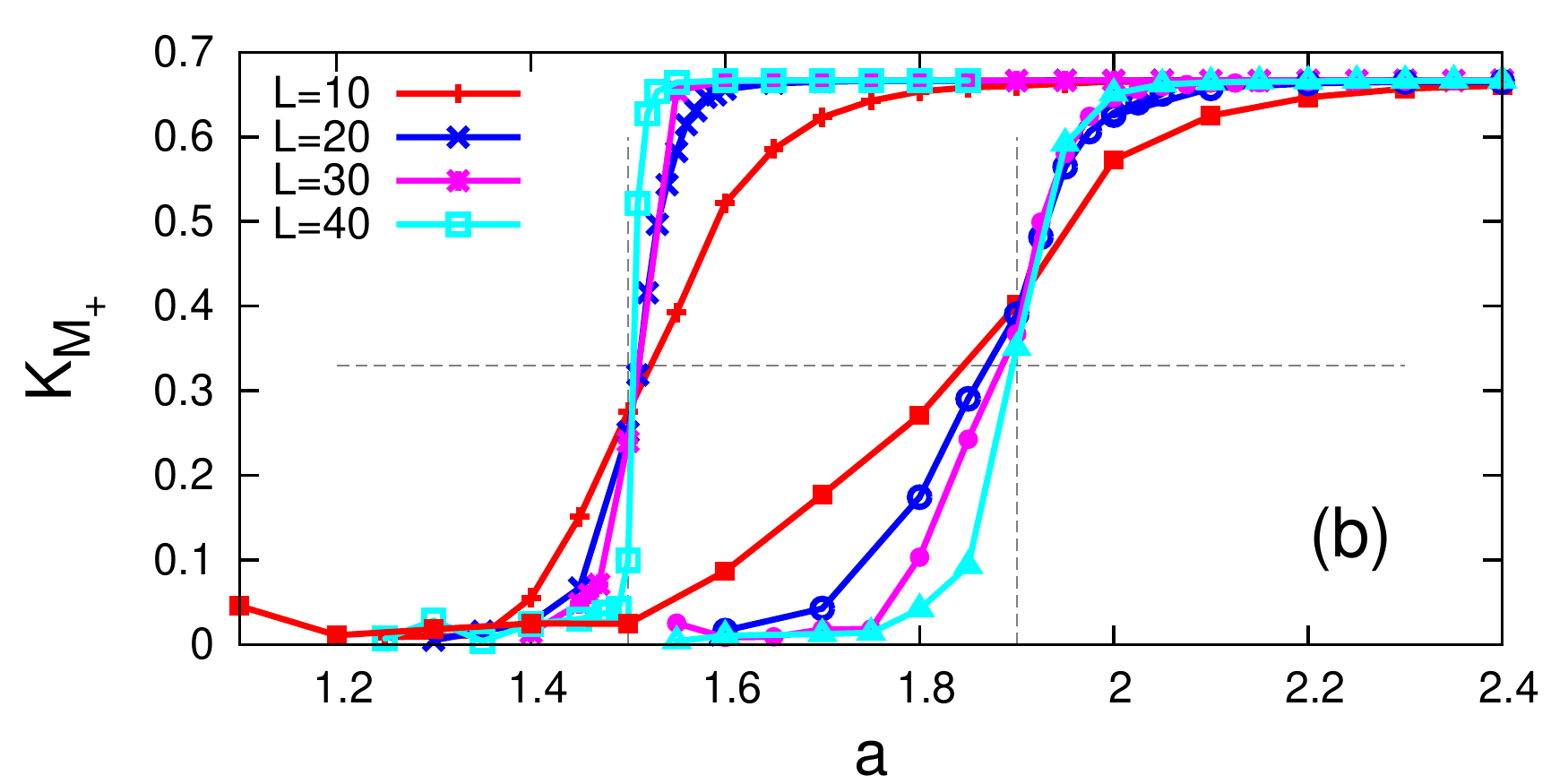} 
\caption{(Color online.)  Study of the FM transition. (a)
  Magnetization per spin $M_+$ and (b) magnetization cumulant
  $K_{M_+}$ as a function of $a$ for $b=0.5$, $d=10^{-5}$ (the group
  of curves on the left) and $d=0.1$ (the ones on the right), and
  several $L$ given in the key. The crossing
  points of $K_{M_+}$ determine $a_c(b,d)$. The vertical dotted (black) lines are
  the critical values predicted by $|\Delta_{16}|=1$. The horizontal dotted level is 
  $1/3$.}
\label{equilibrium}
\end{figure}

The  $2d$ sixteen-vertex model  is isomorphic to the Ising model on 
the checkerboard lattice with many-body interactions~\cite{LiebWuBook}. 
While this model is integrable for a special set of parameters for which the equivalent Ising model 
has two-body interactions only, none of these special cases corresponds to our 
choice of parameters. We then used MC simulations to elucidate the equilibrium phase diagram and 
statistical properties of relatively small samples.
 We set the origin of coordinates on a vertex. 
  The indices  $(\alpha,\beta)$ are the coordinates of each vertex and 
  $((2\alpha+1)/2),\beta)$ and $(\alpha,(2\beta+1)/2)$ locate the mid-points of the
right- and up- pointing bonds. 
We checked equilibration with standard criteria; in particular, we verified
(i)  that the density of vertices stabilizes at long times at the same value starting 
the evolution from very different initial configurations; 
(ii) that the spatially local two-time correlation function depends on the time difference only.
These tests will be shown in~\cite{Levis2011a}.
After verifying equilibration in this way
we computed: (i) The absolute staggered magnetization per spin
defined as $M_{\pm}=(\langle| m_{\pm}^{x}|\rangle + \langle|
m_{\pm}^{y}|\rangle)/2$ with $L^2m^{x}_{\pm}=\sum_{(\alpha,\beta)\in
  A}S_ {(2\alpha+1)/2,\beta}\pm \sum_{(\alpha,\beta)\in
  B}S_{(2\alpha+1)/2,\beta}$ and its counterpart $m^{y}_{\pm}$. 
In the expressions for $m$ we divided the lattice into two sub-lattices $A$ and $B$ such that
$\alpha+\beta$ is even and odd, respectively. The $\pm$ signs allow
one to distinguish between FM and AF order.  (ii) The fourth-order
cumulant $K_{M_{\pm}}=(K_{m_{\pm}^x}+K_{m_{\pm}^y})/2$ with
$K_{m^x_{\pm}}=1-\langle (m^x_{\pm})^4\rangle/3\langle
(m^x_{\pm})^2\rangle$.  $\langle \cdots \rangle$ denotes an average
over independent runs.  In Fig.~\ref{equilibrium}~(a) we show $M_+$
for $b=0.5$ and two values of $d$ as a function of $a$.  $L=10,\dots ,40$
and we averaged over $10^3-10^4$ samples.  The variation of the magnetization as a 
function of $a$ shows a sharp jump at $a=1.5$ for $b=0.5$ and $d=0$, as expected for a first order 
phase transition. For $d>0$ the curve takes a sigmoid form that gets 
wider (less step-like) for increasing $d$. The intersection point appears 
at larger values of $a$ for larger values of $d$. These
features suggest that the transition to the FM phase is continuous,
occurs at larger values of $a$, and that there are fluctuations in the
ordered state. The crossing of
$K_{M_+}$ at height $1/3$ (dotted horizontal line) 
for several values of $L$
shown in Fig.~\ref{equilibrium}~(b) determines $a_c(b,d)$.
Finite size scaling will be discussed in~\cite{Levis2011a}.
Consistently with a second-order phase transition, $K_{M_+}$ remains
positive for all $L$ and the energy cumulant (not shown) converges to
zero.  The analysis of the staggered magnetization and its cumulant
demonstrates that the AF transition no longer belongs to the KT
universality class and also becomes second-order as soon as $d\neq
0$~\cite{Levis2011a}. The projection of the critical lines onto the
$(a,b)$-plane are shown in Fig.~\ref{PhaseDiagram16} with dashed (blue
and black) lines. For $d=10^{-5}$ they agree, within our numerical
accuracy, with the exact critical lines for the six-vertex model. For
increasing $d$ the extent of the PM phase increases. The excitation
properties are radically modified by the relaxation of the ice-rule:
the paramagnetic (PM) phase loses its quasi long-range order and the
FM state admits excitations~\cite{Levis2011a}. These conclusions are in
agreement with exact computations on the eight-~
\cite{LiebWuBook,BaxterBook} and sixteen-~\cite{LiebWuBook}
vertex models for special values of the parameters.  We conjecture
that the phase diagram is characterized by the {\it anisotropy
  parameter}
%\begin{equation}
$ \Delta_{16}=[a^{2}+b^{2}-c^2-(4d)^2]/[2(ab+c(4d))]$.
% \; , 
%\label{Delta16}
%\end{equation}
%where $\tilde{d}=d+4.\frac{3}{4}e$ where $d$ and $e$ are the weight of 2-fold and 1-fold defects respectively. 
%We recover the eight-vertex model anisotropy parameter by setting
%$e=0$. 
The PM phase corresponds to the parameter space sub-manifold
with $|\Delta_{16}|<1$, the FM phase to $\Delta_{16}>1$ and the AF
phase to $\Delta_{16}<-1$ in agreement with the fact that the
transition lines
%$\Delta_{16}=\pm1$,
%i..e. $a^2+b^2-1\mp2ab=(4d)^2\pm2(4d)$, and these 
are parallel to the six-vertex model ones. A similar displacement of the critical lines was
found in spin-ice on the Husimi tree~\cite{Jaubert2009c}.

\begin{figure}[h]
\centering{
\includegraphics[scale=0.28,angle=-90]{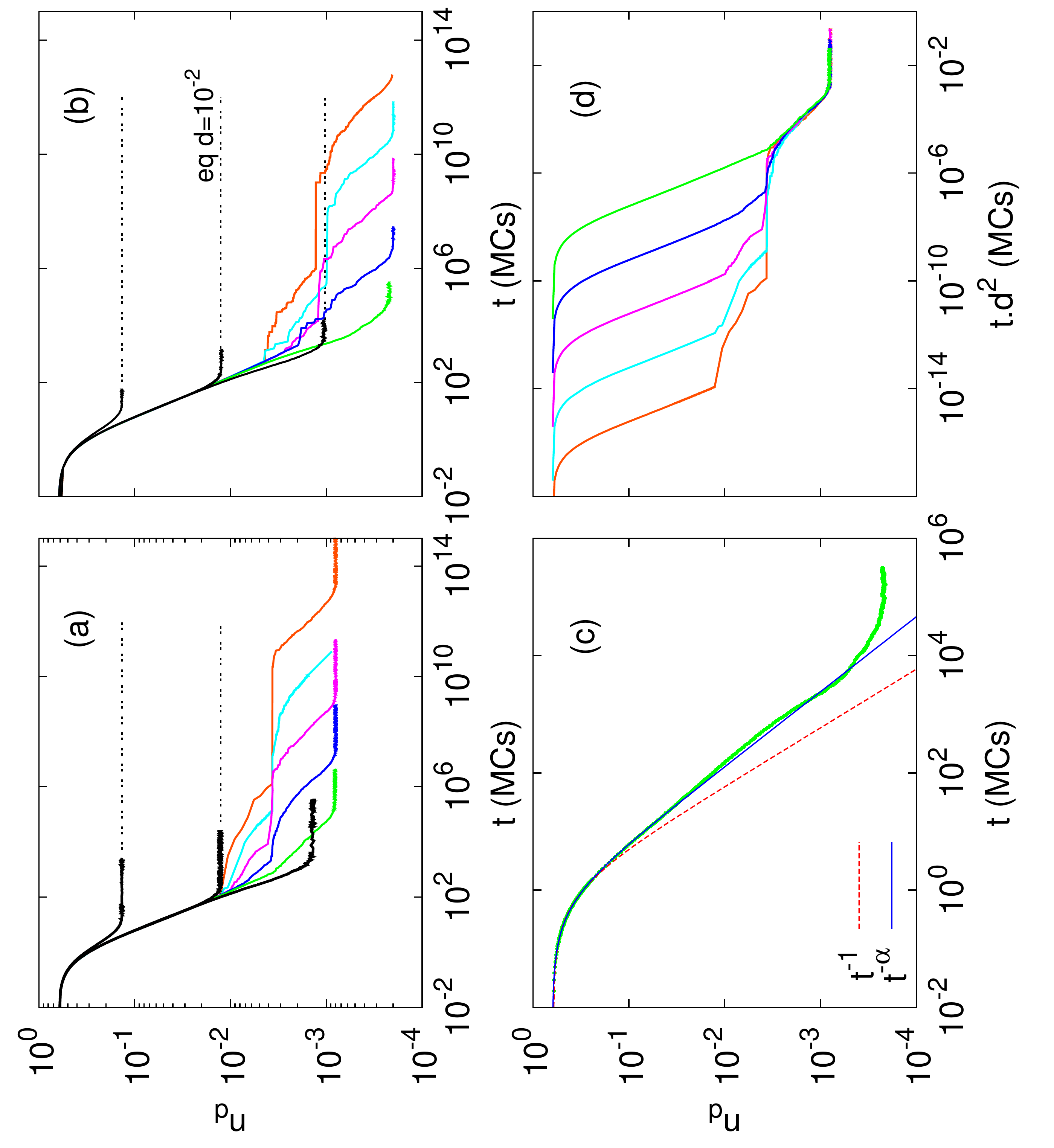}  
%\end{figure}
%\begin{figure}[t]
}
\caption{(Color online.) Time-dependent density of defects, $n_d(t)$,
  after a quench from $T\to\infty$ to $a=b=1$ and $d=10^{-1}$, $10^{-2}, \dots ,$ $10^{-8}$. 
  (a) $L=50$ and (b) $L=100$. The black curves are
  for $d=10^{-1}, \ 10^{-2}, \ 10^{-3}$.  The gray (color) curves are for smaller values of
  $d$ decreasing from left to right.  (c) Short time behavior in the case $d=10^{-4}$ and $L=100$ confronted to the 
  decay $\rho_0/(1+\Omega t)$ (dashed curve)~\cite{Castelnovo2010} 
  and the fit $\rho_0/(1+\Omega t)^\alpha$ with $\alpha=0.78$ (blue plain curve). 
 (d) Test of scaling with $t d^2$ for systems with $L=50$.  }
\label{fig:fit_diffusionprocess}
\label{fig:timescalingd2}
\label{fig:quench-D}
\end{figure}

{\it Quench dynamics.} We now turn to  the dissipative stochastic dynamics of an equilibrium initial
configuration at  $a=b=d=1$ (i.e. $T\to\infty$) after a quench to
sets of parameters in the (i) disordered, (ii) FM, and (iii) AF
phases.  In case (i) the system should equilibrate easily but the
question remains as to whether it gets blocked in metastable states
with a large density of defects.  In cases (ii) and (iii) the
interactions between the spins, mediated by the choice of vertex
weights, should create ordered domains, FM or AF.  The quantitative 
characterization of growth in the ordering processes is given by  two 
possibly different growing lengths extracted from correlation functions 
along 
%\begin{equation}
%L^d_{\parallel,\perp}(t) = \int d^dr\; C_{\parallel,\perp}(\vec r,t)
%\label{eq:growing-lengths}
%\end{equation}
orthogonal directions $\parallel$ and $\perp$ that we identify.

\begin{figure}[h]
\centering
 \includegraphics[scale=0.34]{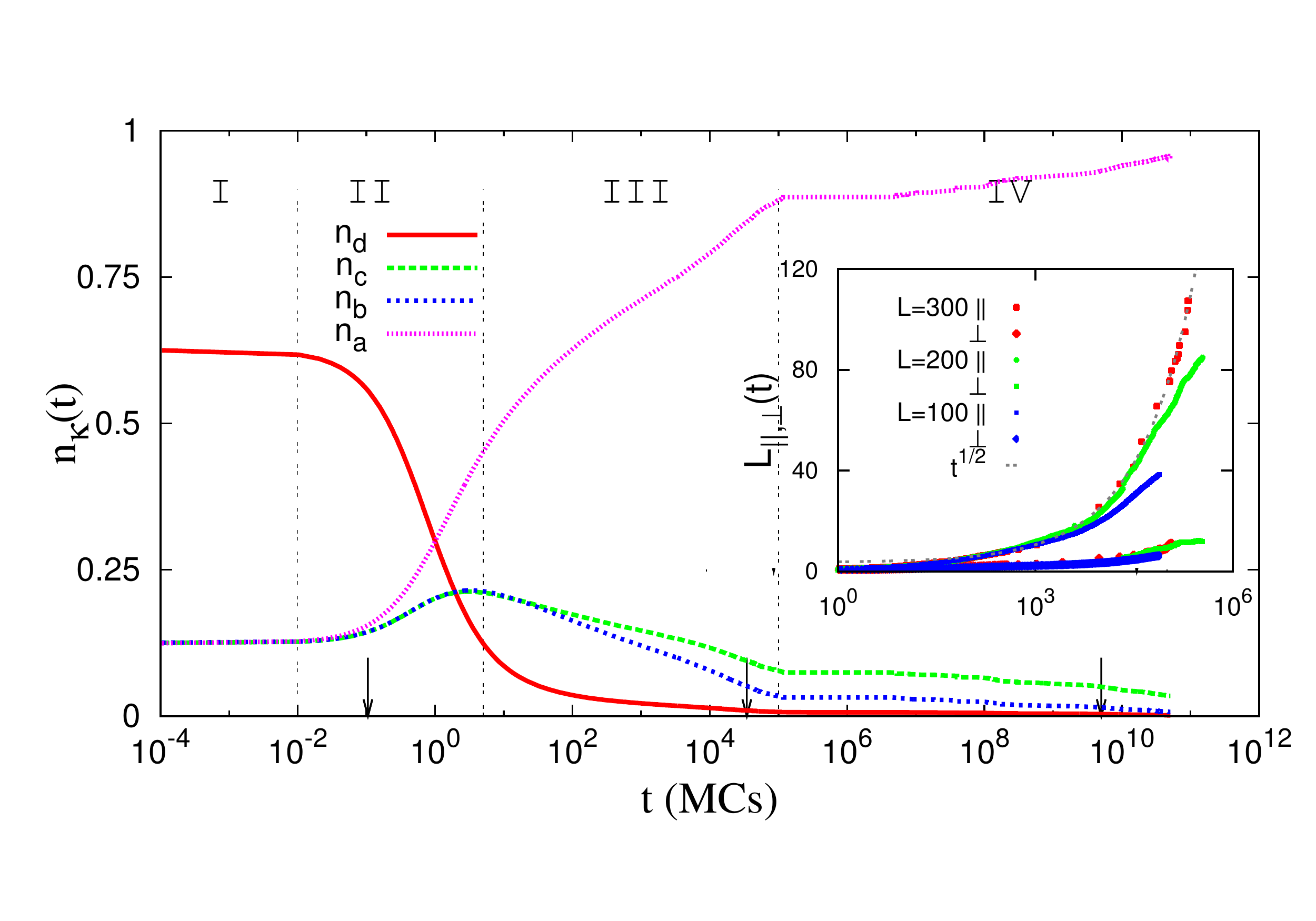}
\par
\ \ \ 
 \includegraphics[width=2.3cm,angle=0]{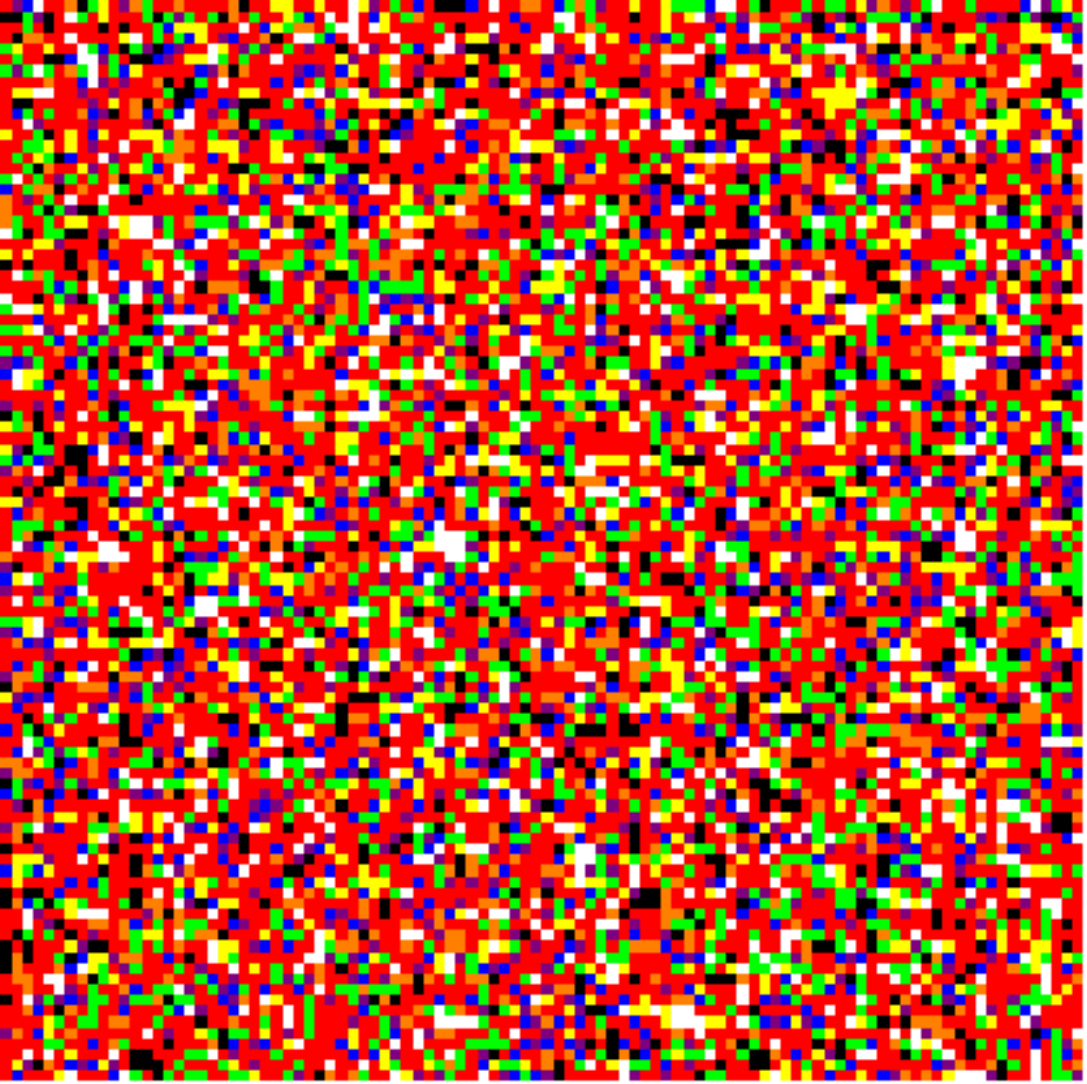}
\ \ \ \    
 \includegraphics[width=2.3cm,angle=0]{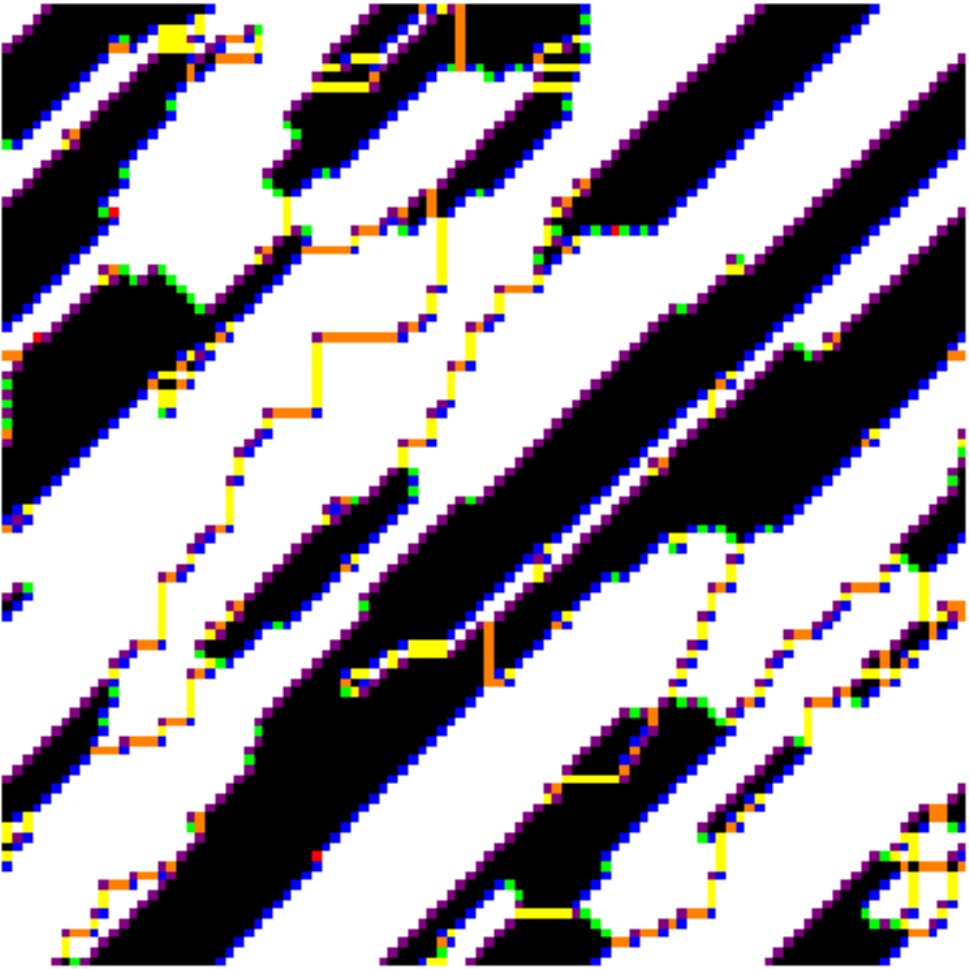}
\ \ \ \    
 \includegraphics[width=2.3cm,angle=0]{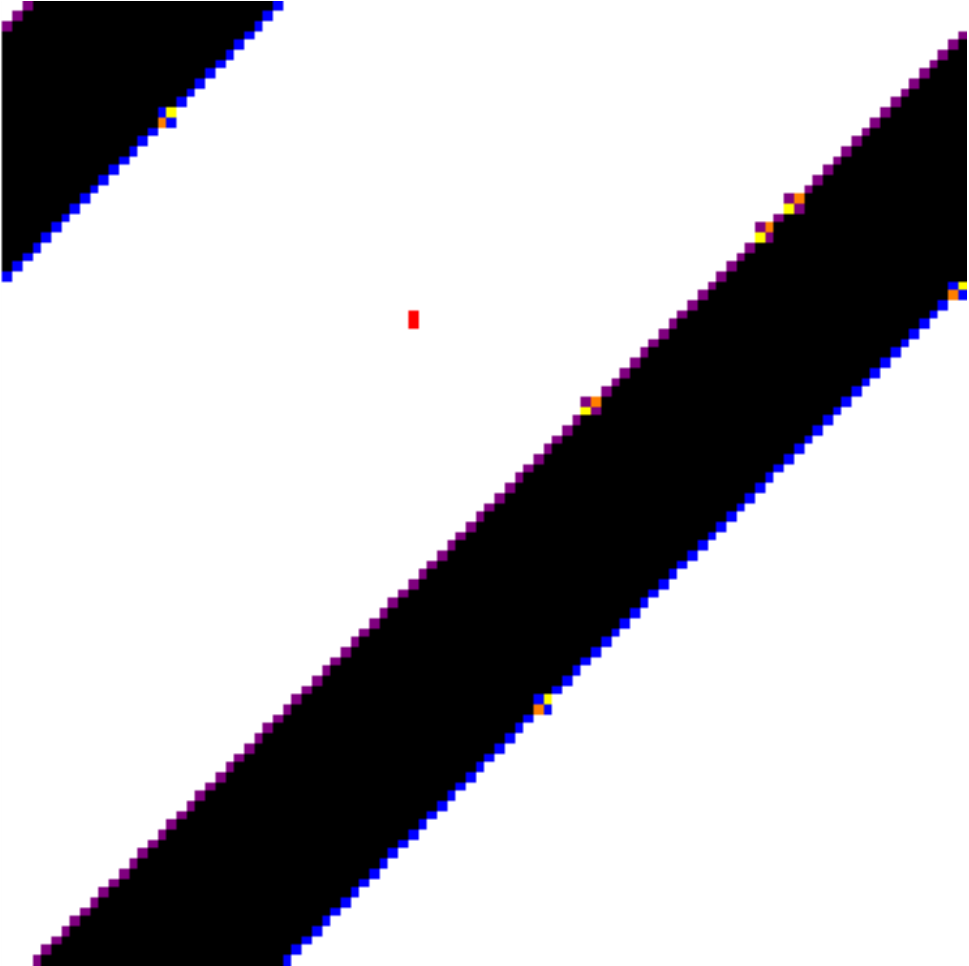}
\caption{(Color online.) FM ordering.  Upper panel: time evolution of
  the density of vertices with weight $a,b,c,d$ for $a=5$, $b=1$,
  $d=10^{-5}$ and $L=100$ averaged over 300 samples.  
  The snapshots are typical configurations
  at the dates indicated by the arrows. Black/white points are vertices 1/2 and the 
  rest are shown in gray (color) scale. Inset:
  time-dependence of the longitudinal (upper curves) and transverse (lower curves) 
  growing lengths for three system sizes,
  $L=100, \ 200, \ 300$. A fit to $t^{1/2}$ is shown with a dotted black line.}
\label{Evolution}
\end{figure}

(i) {\it Quench into the disordered} (D) {\it phase.}  Figure~\ref{fig:quench-D} displays the
time-dependent density of defects, $n_d(t)$, defined as the
number of vertices of type 7-16 divided by $L^2$, 
after an infinitely rapid quench to $a=b=1$ and
$d=10^{-8},\dots ,10^{-1}$ of samples with linear size $L=50$ 
(a) and $L=100$ (b).  
These data have been averaged
over $10^3$ runs.

For large $d$ (black dark curves) 
$n_d(t)$ quickly saturates to its equilibrium value.
Numerical estimates of the equilibrium density of defects, $n_d^{eq}$,
for $d=10^{-1}, \ 10^{-2}, \ 10^{-3}$ are shown with dotted black
lines. As expected $n_d^{eq}$ is an intensive quantity that
increases with $d$. It 
does not depend upon the system size for $L \geq 50$ 
and $d>10^{-3}$.

For small $d$ ($\stackrel{<}{\sim} 10^{-4}$) the systems do not reach
equilibrium within the simulated time-window. After a first decay, $n_d(t)$ gets frozen at
approximately constant values before relaxing, in a much longer
time-scale, to a configuration in which only two defects are present
in our small samples. 
Note that in order to distinguish the $d-$dependent equilibrium
values for these very small $d$s one would need to equilibrate
much larger samples. Unfortunately, the special purpose loop algorithm devised for the 
6 vertex model does not apply to our generalized case. For our working sizes we see asymptotes taking the value $2/L^2$ on average.

The initial decay is fitted by a $\rho_0/(1+\Omega t)^\alpha$
power-law decay with $\alpha \simeq 0.78$  over 
three orders of magnitude in $t$ and $n_d$, 
as shown in panel~(c) in Fig.~\ref{fig:fit_diffusionprocess}. The power-law is shown  with a solid blue line
in the figure together with the data for
$d=10^{-4}$ and $L=100$. 
This law is different from the simple $t^{-1}$ decay found with a mean-field approximation to a 
diffusion-reaction model shown with a dashed red line in the same figure~\cite{Castelnovo2010}.

The decay is next arrested at a metastable density of
defects $n_d^{pl}\approx 10/L^2$. The plateau lasts longer
for smaller $d$ and its  height is roughly independent of $d$. This feature is reminiscent of what was found
 numerically in dipolar spin-ice although contrary to the modeling in~\cite{Castelnovo2010}
our model does not have long-range interactions.
At the entrance to the plateau the system has between 3 and 4 times more defects 
of type 7-8  than those of type 9-16 which are only 2 or 3. 
Therefore, in the final decay from 
the plateau to the asymptotic value $n_d\approx 2/L^2$ the remaining doubly charged defects 
have to disappear. This may be due to  
 two kinds of processes. In the first case, two defects of type 7 and 8 meet to produce two singly (and oppositely)
charged defects with no energetic gain. The total density of defects remains constant after this reaction. 
An example of the second case is a reaction in which 
a defect of type 7 (charge $q=2$) meets one of type 14 (charge $q=-1$) 
to produce a defect of type 10 (charge $q=1$) and a spin-ice vertex with no charge. 
In terms of a reaction-diffusion model this corresponds to 
$2q + (-q) \to q+ 0$ and, consequently, an energetic gain.
Note that the number of single charged defects has not been modified in this process but the 
number of doubly charged defects diminished and so did the total number of defects.
In both cases the remaining defects need to diffuse, a process with no energetic cost, 
to find a partner and annihilate. 
From inspection of the individual runs and the densities of single and doubly charged defects 
we see that the second process is favored, as also suggested by the energetic gain. 
This regime is characterized by a scaling of the dynamic 
curves with the scaling variable $t d^2$~\cite{Castelnovo-priv} as shown in  Fig.~\ref{fig:fit_diffusionprocess}~(d)
for the $L=50$ data.

\begin{figure}[t]
\centering
\onefigure[scale=1]{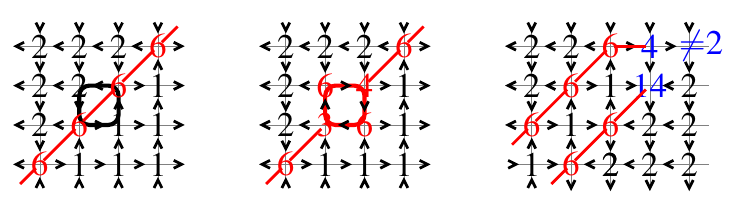} 
\caption{(Color online.) Interfaces between FM domains. Local
  vertices and spins on the bonds are shown. Left panel: diagonal
  wall (red solid line).  Central panel: a `loop' fluctuation 
  on the plaquette highlighted in the left panel. Right
  panel: a $b$ corner vertex cannot be neighbor
  of an $a$-vertex, explaining the presence of strings.}
\label{MecanismosDyn}
\end{figure}
%Mecanismos de desaparicion de bandas.

(ii) {\it Quench into the} FM {\it phase.}  We choose $a=5$, $b=1$ and $d=10^{-5}$,
favoring vertices with weight $a$.  In Fig.~\ref{Evolution} we present
the density of vertices, $n_\kappa(t)$, with $\kappa=a,b,c,d$, 
in a log-linear scale. The evolution is illustrated
with three configurations at instants shown with vertical
arrows. Domains grow anisotropically 
and we choose the $\parallel$ and $\perp$ directions 
to be parallel and perpendicular to the diagonal joining the
lower-left and upper-right corners in the pictures, respectively.
%regimen I
During a short transient ($t\stackrel{<}{\sim} 0.01$ MCs)
all densities remain roughly constant (regime I).
%regimen II
Suddenly, a large number of defects are transformed into
divergence-free vertices by a few single spin-flips: $n_d$ decays
while $n_a$, $n_b$ and $n_c$ increase (regime II) independently  
of $a$ ~\cite{Levis2011a}. A typical
configuration at this stage is the left-most snapshot and there is no 
visual ordering as corroborated by the small values taken by $L_{\parallel,\perp}$ 
and displayed in the inset  in a log-linear scale for three
values of the system size, $L=100$, $L=200$
and $L=300$.
%regimen III
Subsequently the system sets into a slow relaxation regime in which
the dominant mechanism is the one of growing anisotropic domains with
FM order, see the central snapshot (regime III); $n_\kappa$ depend upon
$a$ and there are as many domains with $m^{x,y}=1$ (vertices 1) as
$m^{x,y}=-1$ (vertices 2) respecting symmetry.  
In this regime  $L_\parallel$ grows
faster than $L_\perp$ and tends to saturate 
 to an $L$-dependent value 
when the stripes are fully formed. For the largest sample size, $L=300$, 
our numerical data 
are consistent with a $t^{1/2}$ growth that is shown with a dotted black line. 
Instead order in the $\perp$ direction has not yet percolated. 
The full equilibration of the 
sample needs the percolation of order in the $\perp$ direction which is 
achieved by a still much slower mechanism
(regime IV).

%Mecanismo de movimiento de bandas
\begin{figure}[t]
\onefigure[scale=1]{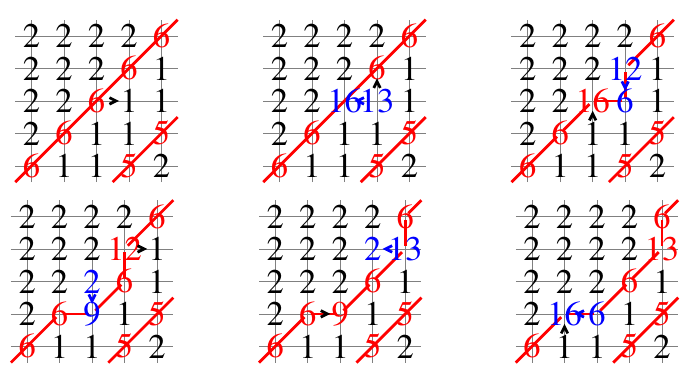} 
\caption{(Color online.) Schematic representation of FM stripe
  motion. Vertices on each site are specified. Diagonal (red) lines
  delimit domains of opposite magnetization. Black arrows indicate the
  spins that flip to get the new configuration (represented in blue after the flip).}
\label{MovimientoBandas}
\end{figure}

A better understanding of the processes involved in the ordering dynamics 
is reached from the analysis of 
the snapshots. (a)
Domain walls are made of
$c$-vertices and plaquettes of divergence-free vertices, as shown in
the left and central panels in Fig.~\ref{MecanismosDyn}, respectively.
The latter are `loop' fluctuations in which all the spins on the
plaquette are sequentially flipped.  Interfaces between FM states tend
to be parallel to the main diagonal, which one depending on
which FM phase one quenches into.  (b) Quasi-one-dimensional
paths made of $b$ and $c$ vertices (loop fluctuation can be attached
to them) act as bridges between two domains of the same type
and run through a region with the opposite order. These structures are
similar to the ones found in the kinetically constrained spiral
model~\cite{Corberi2009}.  In order to further increase the density of $a$-vertices
and develop the FM order the domain
walls and bridges have to be eliminated.  The latter disappear
first via the following mechanism.
%DESCRIPCION DE LOS MECANISMOS
`Corners' made of $b$ (or, less commonly, $d$) vertices sit on a
curved domain wall.  Such $b$ vertices cannot be surrounded by more
than two type 1 or 2 vertices (only defects can, see the third panel
in Fig.~\ref{MecanismosDyn}).  The string progressively disappears
eaten by the attached domains that grow from the corner or,
alternatively, it is first cut by the creation of two defects and the
two strands subsequently shrink, an extremely slow process.  Once the
path has been eliminated one is left with two defects sitting on the
walls of the now detached domains, that move along the interface and
eventually annihilate with their anti-partner. Once parallel bands are
created (third configuration in~Fig.~\ref{Evolution}) the mechanism in
Fig.~\ref{MovimientoBandas} takes over (regime IV). After the creation
of a pair of defects on the interface, the sequence of steps in the
figure  shrink the vertex 1 stripe on a time scale that 
diverges with $L$.

\begin{figure}[h]
\centering
 \includegraphics[scale=0.34]{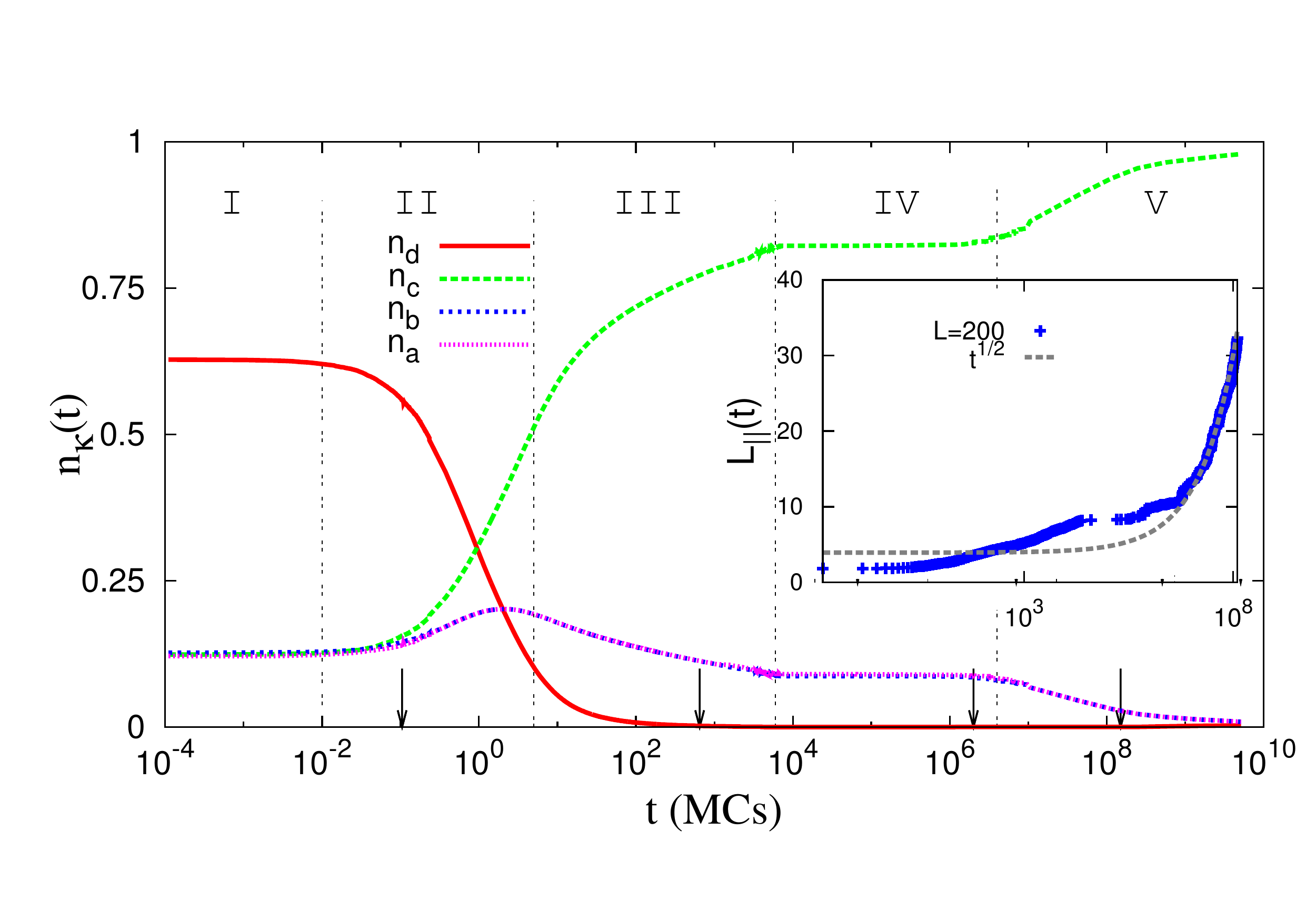}
\par
\ \ 
 \includegraphics[width=1.9cm,angle=0]{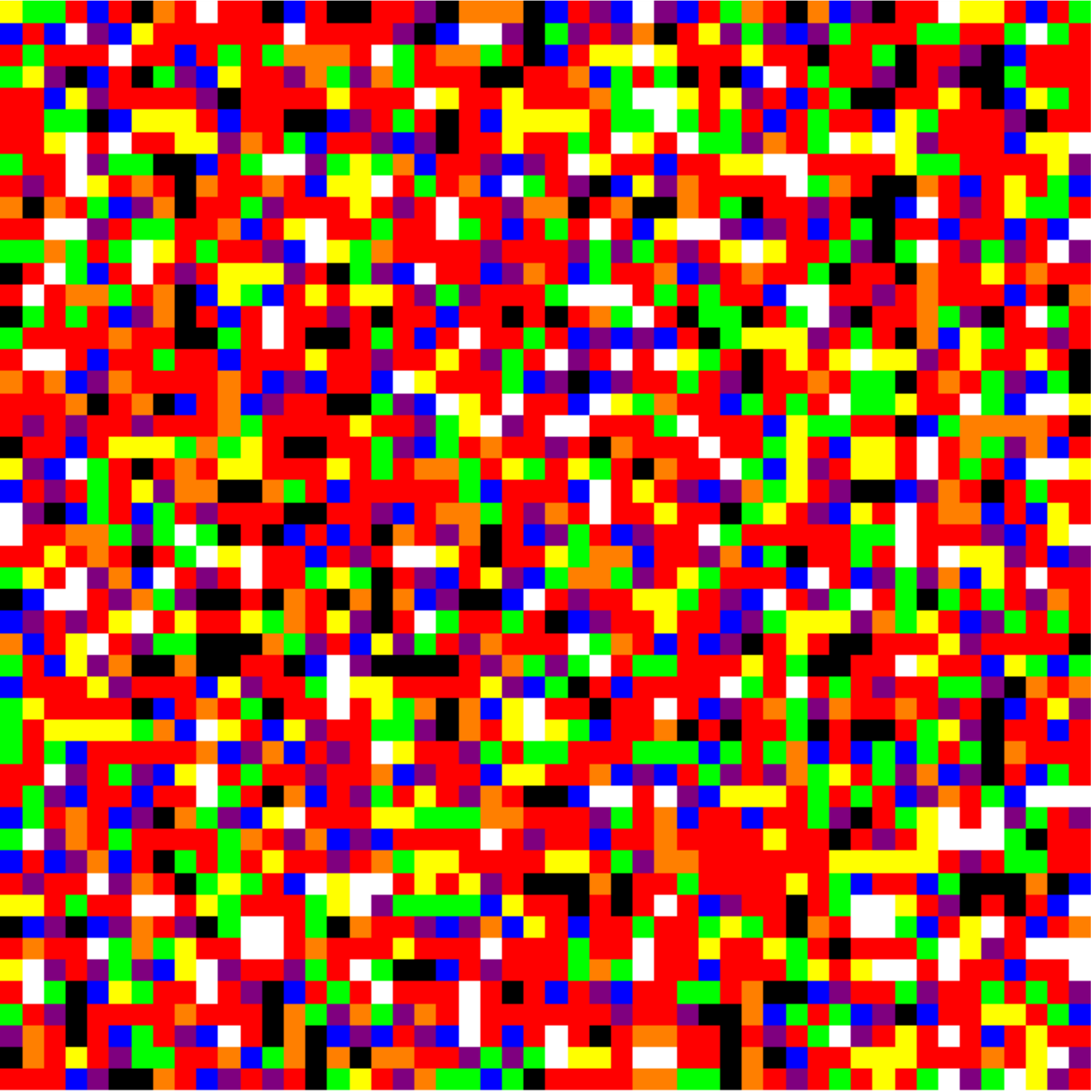}
\ \ 
 \includegraphics[width=1.9cm,angle=0]{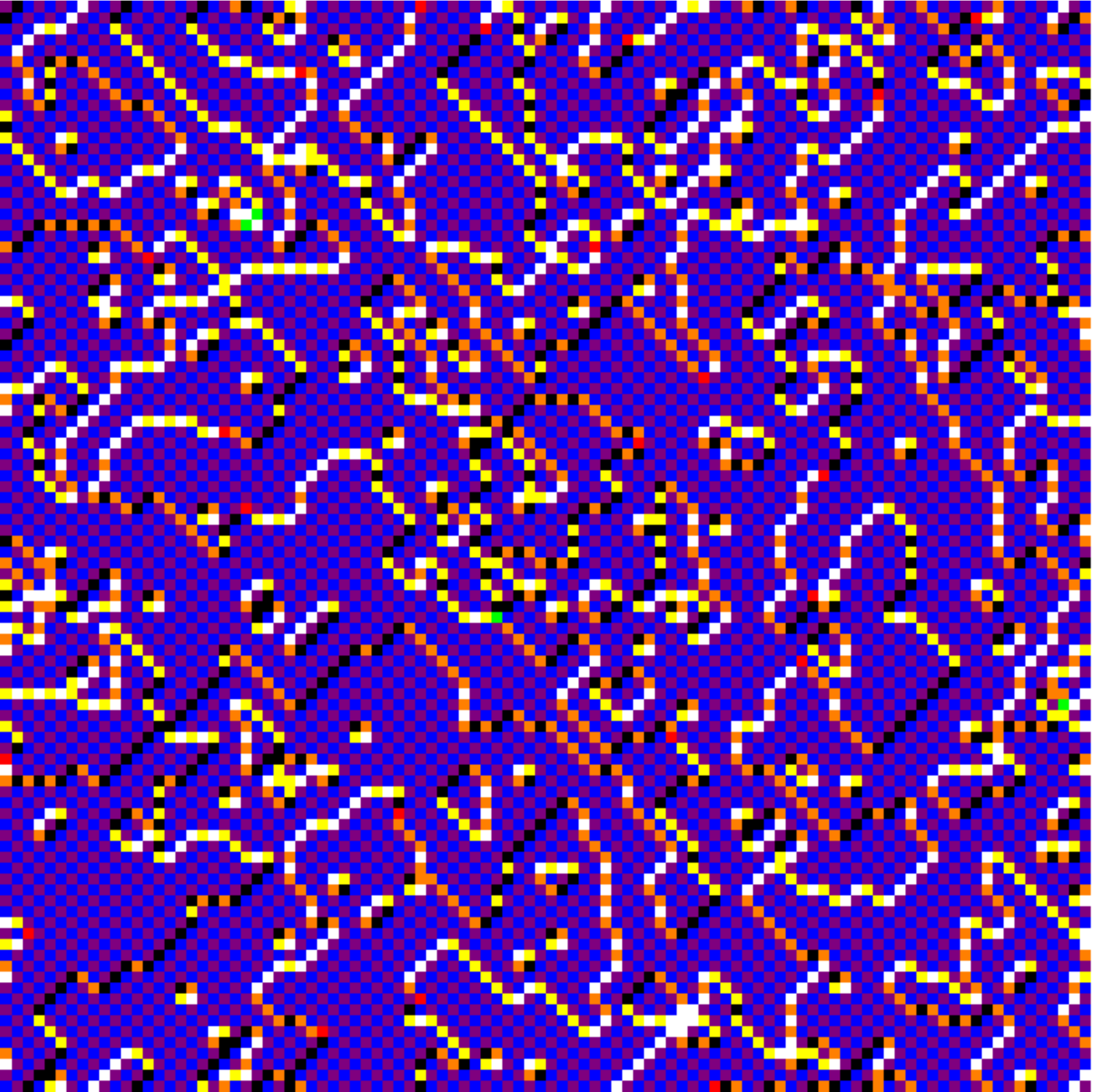}
\ \ 
 \includegraphics[width=1.9cm,angle=0]{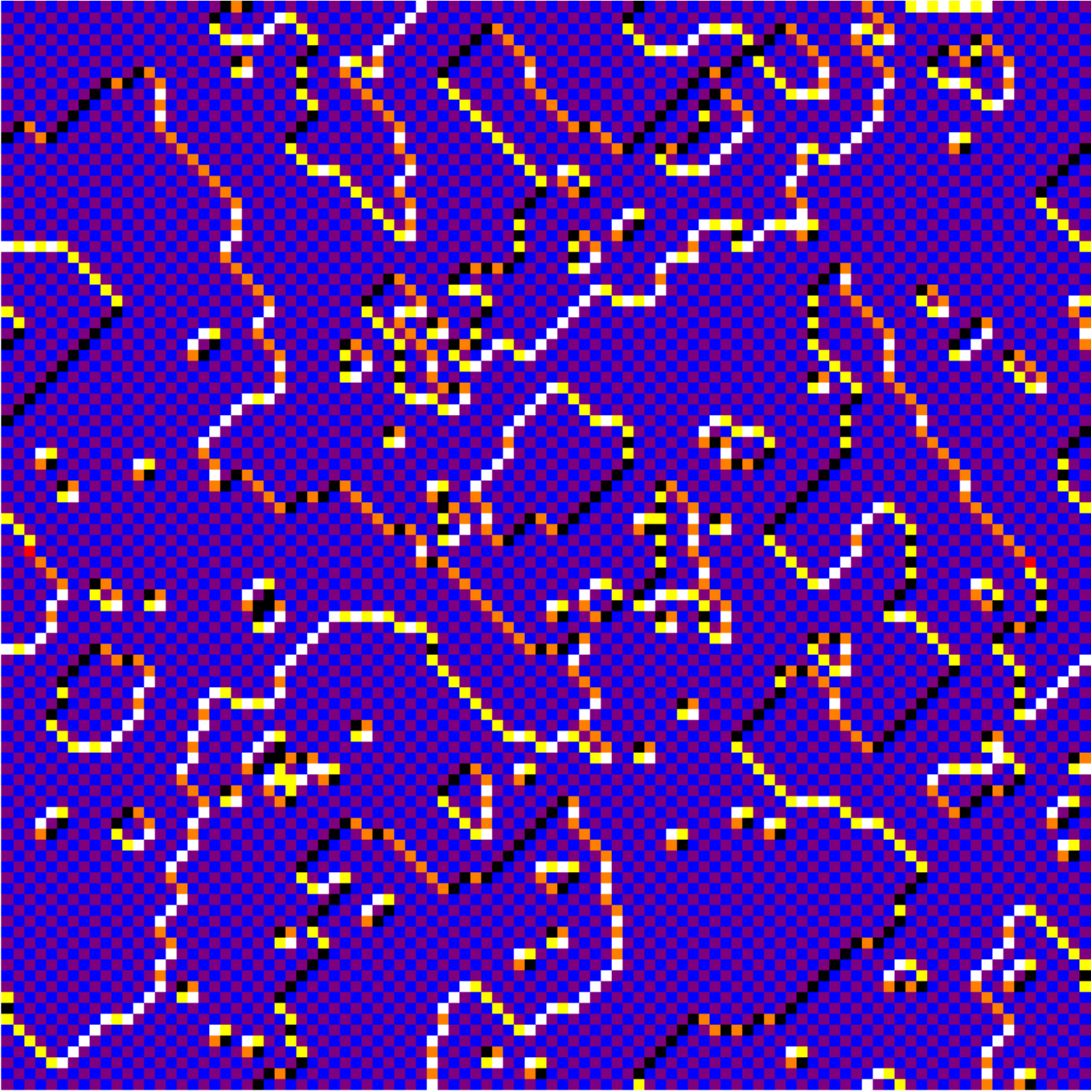}
\ \ 
 \includegraphics[width=1.9cm,angle=0]{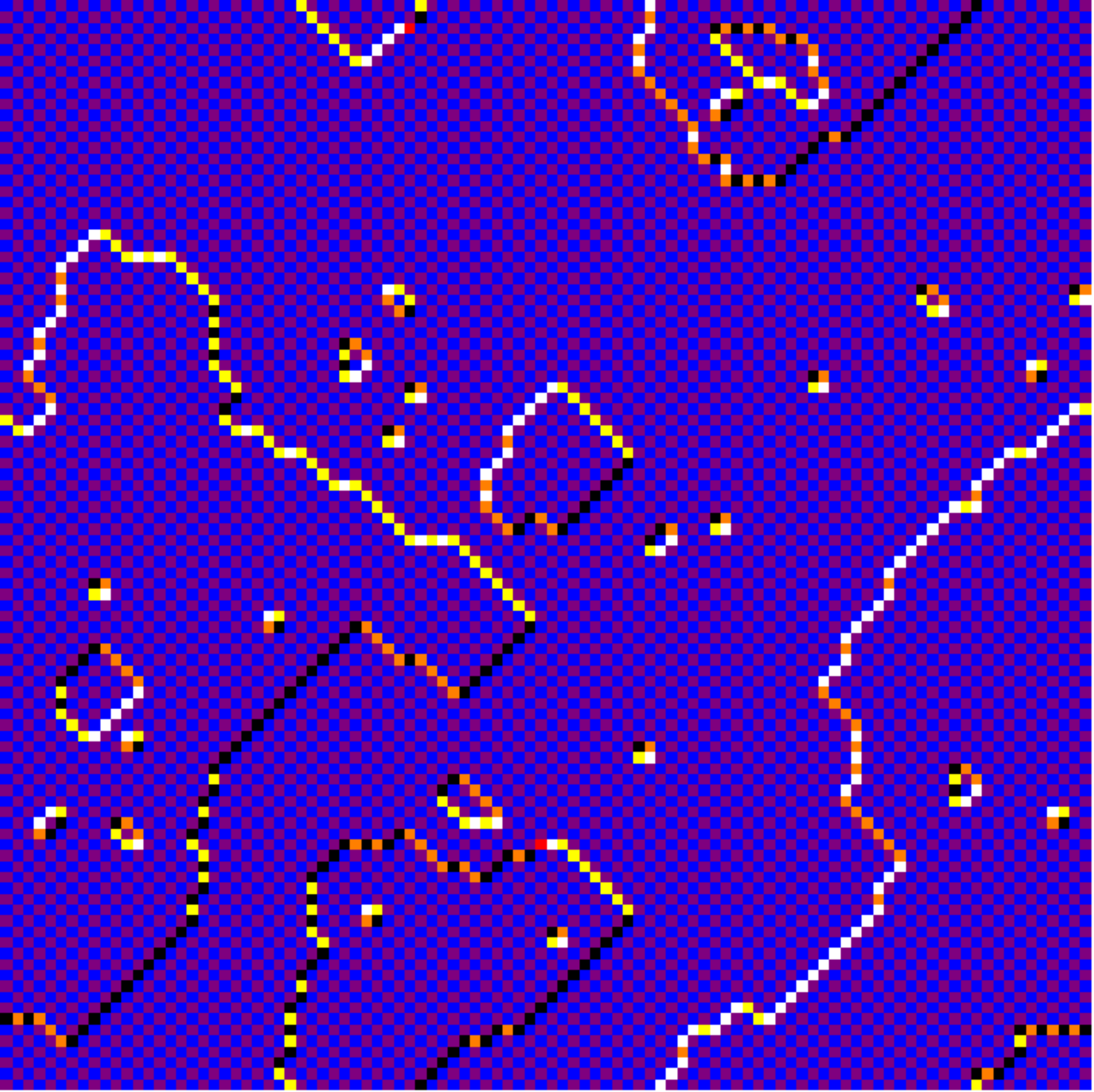}
\caption{(Color online.) AF ordering. Time evolution of the density of vertices in a system with 
$L=100$ after a quench to $a=0.1$, $b=0.1$, $d=10^{-5}$ averaged over 300 runs. Inset: the time-dependent 
growing length $L_\parallel$ 
confronted to $t^{1/2}$ (dotted black line). Typical configurations  are shown.}
\label{EvolutionAF}
\end{figure}

(iii) {\it Quench into the} AF {\it phase.}  The evolution of the vertex population is
shown in the main panel in Fig.~\ref{EvolutionAF} for $a=b=0.1$ and
$d=10^{-5}$. Similarly to what found in the FM quenches, in regime I
all densities remain approximately constant. This is followed by
regime II with a rapid annihilation of defects into divergence-free
vertices. The creation of $a, \ b$ and $c$-vertices occurs with a rate
that depends on $a$ while, surprisingly, $n_d$ does not, at least within 
our numerical accuracy.  In regime
III the system increases the AF order by growing domains of staggered
magnetization $\pm1$ with $c$ vertices. Since $a$ is very close to $b$
for our choice of parameters, domains 
are quite isotropic and the growing length are, within numerical accuracy, 
$t^{1/2}$. Regime IV follows next and it is
characterized by a strong slowing-down although there is 
no obvious extended structure blocking the evolution. 
In regime V the system finally
reaches equilibrium. The relevant elementary moves in each regime will be discussed
in~\cite{Levis2011a}.

We presented a numerical study of the quench dynamics of
$2d$ spin-ice. In this Letter we established the main statics and dynamical properties of this model: equilibrium phase diagram and relaxation dynamics after thermal quenches. No study of frustrated magnet existed in the litterature such complete. We demonstrated the existence of long-lived metastable
states with an excess of topological defects in a model with no
long-range interactions, cfr.~\cite{Castelnovo2010}. We
showed that the dynamics after quenches into the FM and AF phases 
proceed by coarsening of equilibrium domains. More details 
will be presented in an extended publication~\cite{Levis2011a}.
Our results should motivate new experimental studies of the 
dynamics in frustrated magnets and possibly motivate researchers in the integrable systems community to try to adapt their very powerful techniques to deal with dynamical issues. 

\acknowledgments
We thank R. Borzi, C. Castelnovo, 
L. D. C. Jaubert and S. Grigera for useful discussions. This work was 
financially supported by ANR-BLAN- 0346 
(FAMOUS).
 
\bibliographystyle{eplbib}
\bibliography{EPLshort}

\end{document}